\documentclass[conference]{IEEEtran}
\usepackage{amsmath,amsthm}
\usepackage{cite}
\usepackage{graphicx}
\usepackage{epstopdf}
\usepackage{amsfonts,amsmath,amssymb}
\usepackage{cite}
\usepackage{graphicx}
\usepackage{url}
\usepackage{bm}
\usepackage{bbm}
\usepackage{amssymb}

\begin{document}

\title{{CV-QKD with Gaussian and non-Gaussian Entangled States over Satellite-based Channels}}
\author{
\IEEEauthorblockN{Nedasadat Hosseinidehaj and Robert Malaney}
\IEEEauthorblockA{School of Electrical Engineering  \& Telecommunications,\\
The University of New South Wales,\\
Sydney, NSW 2052, Australia\\
neda.hosseini@unsw.edu.au, r.malaney@unsw.edu.au}
}

\vspace{-5cm}

\maketitle
\begin{abstract}
In this work we investigate the effectiveness of continuous-variable (CV) entangled states, transferred through high-loss atmospheric channels, as a means of viable quantum key distribution (QKD) between terrestrial stations and low-Earth orbit (LEO) satellites. In particular, we investigate the role played by the Gaussian CV states as compared to non-Gaussian states. We find that beam-wandering induced atmospheric losses lead to QKD performance levels that are in general quite different from those found in fixed-attenuation channels. For example, circumstances can be found where no QKD is viable at some fixed loss in fiber but is viable at the same mean loss in fading channels. We also find that, in some circumstances, the QKD relative performance of Gaussian and non-Gaussian states can in atmospheric channels be the reverse of that found in fixed-attenuation channels. These findings show that the nature of the atmospheric channel can have a large impact on the QKD performance. Our results should prove useful for emerging global quantum communications that use LEO satellites as communication relays.
\end{abstract}

\section{Introduction}
QKD allows two distant and trusted parties, Alice and Bob, to create a secret key through the use of quantum and classical communications, and  can be implemented using discrete-variable (DV) or CV quantum states. In DV-QKD protocols, key information is encoded on the properties of single photons, and in CV-QKD protocols key information is encoded on the quadrature variables of  light. In the former technology detection is realized by single-photon detectors e.g., \cite{DV1, DV2, DV3}, while in the latter technology detection is realised by  more efficient and faster detectors, such as homodyne (or heterodyne) detectors e.g., \cite{CV1, CV2, thesis, Weedbrook2012, Weedbrook2013}.

QKD is mostly implemented experimentally in a prepare-and-measure (PM) type scheme e.g., \cite{exp-PM}, where Alice prepares the quantum states which are then transmitted over an insecure quantum channel towards Bob, who measures the received quantum states, resulting in creation of  correlated data between the two trusted parties. Each PM scheme can be represented in an equivalent entanglement-based (EB) scheme, where Alice generates a two-mode entangled state, with one mode being kept by Alice and the other mode being transmitted over an insecure quantum channel. Alice and Bob then proceed to invoke a QKD protocol by measuring their own modes to create  correlated data. In both the PM scheme and EB scheme, after generation of the correlated data, Alice and Bob proceed  with classical postprocessing including sifting, parameter estimation, reconciliation and privacy amplification over a public (but authenticated) classical channel, to generate a secret key even in the presence of Eve (a potential eavesdropper).

CV-QKD protocols using Gaussian resources have been well analysed in theory and experimentally implemented e.g., \cite{thesis, Weedbrook2012, Weedbrook2013, exp-PM}. In the Gaussian PM scheme \cite{thesis, Weedbrook2012, exp-PM}, the prepared quantum states are either coherent states or squeezed states which are modulated by  Gaussian distributions. Each Gaussian PM scheme can be represented by an equivalent EB scheme \cite{thesis, Weedbrook2012, Weedbrook2013} in which Alice generates a two-mode squeezed vacuum (TMSV) state. Although Gaussian quantum states are a well-established resource from both a theoretical and an experimental perspective (for review see\cite{Weedbrook2012}), the use of non-Gaussian states in the implementation of CV-QKD protocols has also garnered interest, e.g. \cite{nG-modulation, nG1, nG2, nG-coherent}.
Such studies can be motivated in part  by the fact that non-Gaussian operations such as photon subtraction \cite{1st_PSS, 2, telep, 3, 9, Oxford, Oxford2013, 7} and photon addition \cite{1st_PAS, telep, Oxford, 7, added} on an incoming TMSV state can lead to higher levels of entanglement  \cite{Oxford}.

It is widely anticipated that the use of satellites will assist in the  deployment of QKD  over global scales. Thus, it is important to analyse the effectiveness of CV-QKD protocols over atmospheric channels towards (and from) LEO satellites. Such channels are highly fading in nature since the transmissivity of the channel fluctuates due to atmospheric effects. CV-QKD protocols using  Gaussian resources have been studied over atmospheric channels in \cite{Usenko, Heim, Neda1, Neda2}. However,  works on using  non-Gaussian resources in CV-QKD protocols have focussed on fixed-attenuation channels. Thus, it remains unclear whether non-Gaussian entangled states can be effective in the implementation of CV-QKD protocols over  atmospheric fading channels (in terms of the quantum key generation rates).  The entanglement-generation rate produced by non-Gaussian entangled states passing through atmospheric channels has been recently studied by us \cite{Neda3}. In this present work we extend such studies to consider bounds on the actual quantum key rates generated by such non-Gaussian transfer. We will be particularly interested in comparing such quantum key rates with those arising from Gaussian transfer over atmospheric channels, and Gaussian and non-Gaussian transfer over fixed-attenuation channels.

\begin{figure}[!t]
    \begin{center}
   {\includegraphics[width=3.4 in, height=2.5 in]{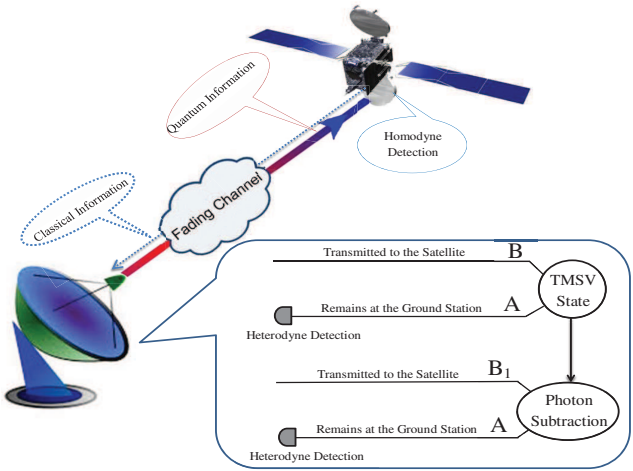}}
    \caption{An example of quantum communication scheme in the ground-to-LEO configuration for implementation of our CV-QKD protocol.}\label{scheme1}
    \end{center}
\end{figure}

We will focus on transmission fading caused by the  beam wander \cite{Usenko, fso, beamwander}, which is expected to dominate photon losses in Earth-to-satellite channels. To explore  non-Gaussian key generation  we will utilize non-Gaussian entangled states which are created just-in-time via photonic subtraction from  incoming Gaussian states. In our security analysis, we will include the effects of channel fading, the probabilistic production of the non-Gaussian states, and the most likely imperfections in a CV-QKD protocol (excess noise, inefficient and noisy detectors, and non-perfect reconciliation algorithms). A schematic illustration of the CV-QKD protocol we adopt in a fading uplink configuration is shown in Fig.~\ref{scheme1}.

The structure of the remainder of this paper is as follows. In Section II, our CV-QKD protocol using a non-Gaussian resource is described, and its security analysis is discussed in detail. In Section III, the performance of the protocol over the atmospheric channels is presented and compared to the corresponding protocol over the fixed-attenuation channels. Finally, concluding results are provided in Section IV.

\section{system model and quantum key rate}
 We now describe the implementation of a CV-QKD protocol using non-Gaussian states and outline how to determine the quantum key rates of this protocol for the atmospheric channel.

\subsection{CV-QKD Protocol using non-Gaussian states}


In this work, we will focus on the implementation of a specific CV-QKD protocol in the EB scheme, in which a two-mode non-Gaussian entangled state is shared between Alice and Bob. Initially, we focus on the case where the non-Gaussian state is created just-in-time via photonic subtraction from an incoming Gaussian state - a non-Gaussian operation which has previously been experimentally demonstrated \cite{exp1, exp2}.

The protocol begins with Alice possessing a TMSV state, $\rho_{AB}$, described in the Fock basis as ${\left| {TMSV} \right\rangle _{AB}} = \sum\limits_{n = 0}^\infty  {{q_n}} {\left| n \right\rangle _A}{\left| n \right\rangle _B}$, where ${q_n} = {\lambda ^n}\sqrt {1 - {\lambda ^2}} $, and where $\lambda  = \tanh r $ with $r$ being the squeezing parameter (indices $A$ and $B$ indicate the two modes). The two-mode squeezing in dB is given by $ - 10{\log _{10}}\left( {\exp ( - 2r)} \right)$. Since the TMSV state is a Gaussian state, it can be completely described by its first moment, which is zero, and its covariance matrix (CM) given by
\begin{eqnarray}\label{TMSV-CM}
M_{AB}^{in}= \left( {\begin{array}{*{20}{c}}
{v\,I}&{\sqrt {{v^2} - 1} \,Z}\\
{\sqrt {{v^2} - 1} \,Z}&{v\,I}
\end{array}} \right) ,
\end{eqnarray}
where $I$ is a $2 \times 2$ identity matrix, $Z = diag\left( {1, - 1} \right)$, and $v = \cosh \left( {2r} \right)$ is the quadrature variance of each mode.
While mode~$A$ is held by Alice, the other mode undergoes the  photon-subtraction operation. To invoke this operation, Mode~$B$ of Alice's TMSV state interacts with a vacuum mode in a beam splitter BS1 with transmissivity $T$ and reflectivity $1-T$, with one of the outputs of the beam splitter feeding a single-photon detector. When the detector registers one photon, a pure photon-subtracted squeezed (PSS) state is heralded in mode~$B_1$, the non-measured output of the beam splitter, defined as ${\rho _{A{B_1}}}$, which is a non-Gaussian entangled state. The normalized state arising from this process and its creation probability $P_{ss}$ are given by ${\left| {PSS} \right\rangle _{A{B_1}}} = \sum\limits_{n = 0}^\infty  {{q_n}} {\left| {n + 1} \right\rangle _A}{\left| n \right\rangle _{B_1}}$, where \cite{Neda3, Oxford}
\begin{eqnarray}\label{PSS}
\begin{array}{l}
{q_n} = \left( {1 - {\lambda ^2}T} \right){\left( {\lambda \sqrt T } \right)^n}\sqrt {n + 1}, \\
\\
{P_{ss}} = \frac{{{\lambda ^2}\left( {1 - {\lambda ^2}} \right)\left( {1 - T} \right)}}{{{{\left( {1 - {\lambda ^2}T} \right)}^2}}}.
\end{array}
\end{eqnarray}
After applying the photon subtraction, the output of the beam splitter, mode~$B_{1}$ is transmitted towards Bob over an unsecured quantum channel with transmissivity $\tau$. In any real-world implementation of this CV-QKD protocol, the quantum state preparation at Alice's side will produce some additional noise. In the following we refer to this excess noise as $\varepsilon$.

The QKD protocol then proceeds with Bob making a homodyne detection of the amplitude or phase quadrature of the received mode $B_2$, and Alice measures both quadratures using a heterodyne detection on mode~$A$ to create correlated data.

\begin{figure}[!t]
    \begin{center}
   {\includegraphics[width=3.4 in, height=2 in]{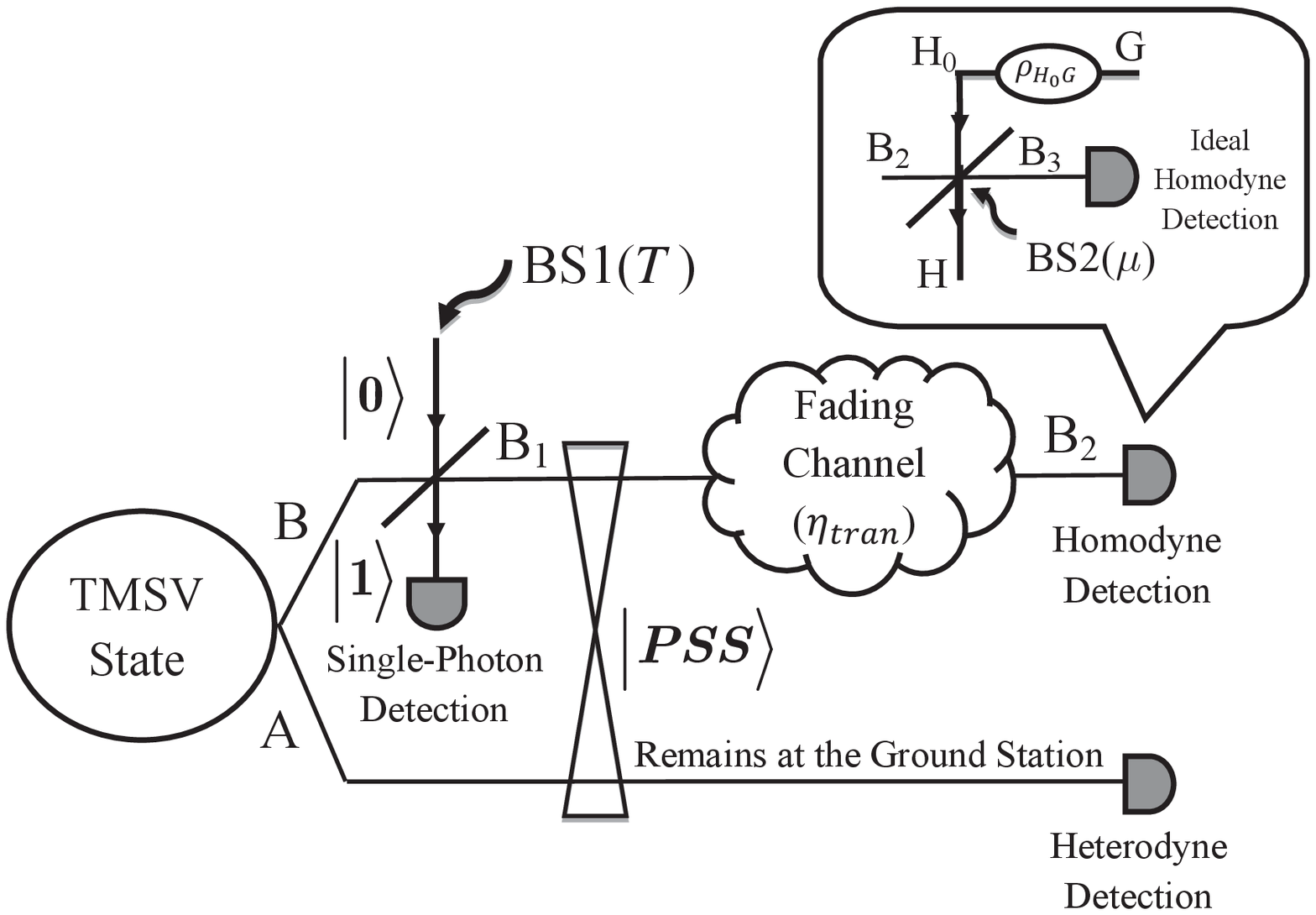}}
    \caption{The entanglement-based CV-QKD protocol over fading channels using the PSS state which is created just-in-time from the incoming TMSV state.}\label{scheme2}
    \end{center}
\end{figure}

Bob's realistic  homodyne detection will have an efficiency $\mu$ and an electronic noise $\nu_{el}$ \cite{thesis, inefficient_homodyne}. The efficiency can be \emph{modeled} by placing a beam splitter BS2 of transmissivity $\mu$ before an ideal homodyne detector. This detector's electronic noise can be modeled by an EPR state, $\rho_{{H_0}G}$, of quadrature variance $\nu _d$, where ${\nu _{el}} = \left( {1 - \mu } \right)\left( {{\nu _d} - 1} \right)$. One input port of the beam splitter BS2 is the received mode $B_2$, and the second input port is fed by one half of the EPR state, mode~$H_0$, while the output ports are mode~$B_{3}$ (which is measured by the ideal homodyne detector) and mode $H$. Our CV-QKD protocol using the PSS state is shown in Fig.~\ref{scheme2}.

Since non-Gaussian states are not completely described by the first and second moments of the quadrature operators, we are not able to quantify the evolution of our PSS state solely through the CM. Thus, we will employ the Kraus representation \cite{Kraus} in order to analyze the evolution of our PSS state through the channel. Considering a quantum state with density operator ${\rho _{in}}$ as the input of a trace-preserving completely positive channel, the output density operator of the channel can be described in an operator-sum representation of the form
${\rho _{out}} = \sum\limits_{\ell  = 0}^\infty  {{G_\ell }{\rho _{in}}\,G_\ell ^\dag }$,
where the Kraus operators ${G_\ell }$ satisfy $\sum\limits_{\ell  = 0}^\infty  {{G_\ell }\,G_\ell ^\dag }  = I$, with $I$ being the identity operator.

The Kraus operators of a fixed-attenuation channel with transmissivity $\tau  $ can be written as \cite{Kraus},
\begin{eqnarray}\label{Atten}
{G_\ell }\left( \tau  \right) = \sum\limits_{m = 0}^\infty  {\sqrt {{}^{m + \ell }{C_\ell }} {{\left( {\sqrt {1 - \tau } } \right)}^\ell }{{\left( {\sqrt \tau  } \right)}^m}\left| m \right\rangle \left\langle {m + \ell } \right|,}
\end{eqnarray}
where ${}^{m + \ell }{C_\ell }$ is the binomial coefficient. From these operators it can then be shown that the elementary density operator $\left| m \right\rangle \left\langle n \right|$ after the evolution through the fixed-attenuation channel can be written as \cite{Kraus},
\begin{equation}\
\begin{array}{l}
\left| m \right\rangle \left\langle n \right| \to \sum\limits_{\ell  = 0}^\infty  {{G_\ell }\left( \tau  \right)\left| m \right\rangle \left\langle n \right|\,G_\ell ^\dag } \left( \tau  \right) = \\
\\
\sum\limits_{\ell  = 0}^{\min \{ m,n\} } {\sqrt {{}^m{C_\ell }{}^n{C_\ell }} } {\left( {1 - \tau } \right)^\ell }{\left( {\sqrt \tau  } \right)^{\left( {m + n - 2\ell } \right)}}\left| {m - \ell } \right\rangle \left\langle {n - \ell } \right|.
\end{array}\label{Evol}
\end{equation}
Considering the initial density operator of our PSS state as the following
\begin{eqnarray}\label{initial}
{\rho _{A{B_1}}} = \sum\limits_{m = 0}^\infty  {\sum\limits_{n = 0}^\infty  {{q_m}} {q_n}{{\left| {m + 1} \right\rangle }_A}{{\left\langle {n + 1} \right|}_A} \otimes {{\left| m \right\rangle }_{{B_1}}}{{\left\langle n \right|}_{{B_1}}}},
\end{eqnarray}
the density operator of the output mixed state ${\rho _{A{B_2}}}$ can be calculated in the Fock basis through the use of Eq.~\eqref{Evol}, giving
\begin{equation}\label{output}
{\rho _{A{B_2}}} = \sum\limits_{a' = 1}^\infty  {\sum\limits_{b' = 0}^\infty  {\sum\limits_{c' = 1}^\infty  {\sum\limits_{d' = 0}^\infty  {{\rho _{a'b'c'd'}}{{\left| {a'} \right\rangle }_A}{{\left\langle {c'} \right|}_A} \otimes {{\left| {b'} \right\rangle }_{{B_2}}}{{\left\langle {d'} \right|}_{{B_2}}}} } } } ,
\end{equation}
with
\begin{equation}\label{elements}
\begin{array}{l}
{\rho _{a'b'c'd'}} = {q_{a' - 1}}\,{q_{c' - 1}} \times \\
\\
\,\,\,\,\,\,\,\,\,\,\,\,\,\,\sqrt {{}^{a' - 1}{C_{a' - 1 - b'}}{}^{c' - 1}{C_{c' - 1 - d'}}} {\left( {1 - \tau } \right)^{a' - 1 - b'}}{\left( {\sqrt \tau  } \right)^{b' + d'}}
\end{array}
\end{equation}
if $a' - b' = c' - d' $ and $a' - 1 - b' \ge 0$, otherwise $\rho _{a'b'c'd'} = 0$.

We can compute the lower bound on the actual key rate, $K$, of the protocol by replacing the shared non-Gaussian state between Alice and Bob $\rho_{A{B_2}}$ with a Gaussian state $\rho _{A{B_2}}^G$ having the same mean value and CM, resulting in $K(\rho _{A{B_2}}^G) \le K({\rho _{A{B_2}}})$. The output mixed state ${\rho _{A{B_2}}}$ has zero mean and CM in the following form
\begin{equation}\label{AB2}
{M_{A{B_2}}} = \left( {\begin{array}{*{20}{c}}
{x'I}&{z'Z}\\
{z'Z}&{y'I}
\end{array}} \right),
\end{equation}
where
\begin{equation}\label{x'y'z'}
\begin{array}{l}
x' = \sum\limits_{a' = 1}^\infty  {\sum\limits_{b' = 0}^\infty  {\sum\limits_{c' = 1}^\infty  {\sum\limits_{d' = 0}^\infty  {(2a' + 1){\rho _{a'b'c'd'}}} } } } \left| {_{a' = c'}} \right.,\\
\\
y' = \sum\limits_{a' = 1}^\infty  {\sum\limits_{b' = 0}^\infty  {\sum\limits_{c' = 1}^\infty  {\sum\limits_{d' = 0}^\infty  {(2b' + 1){\rho _{a'b'c'd'}}} } } } \left| {_{a' = c'}} \right.,\\
\\
z' = \sum\limits_{a' = 1}^\infty  {\sum\limits_{b' = 0}^\infty  {\sum\limits_{c' = 1}^\infty  {\sum\limits_{d' = 0}^\infty  {\left( {\sqrt {a'b'} {\rho _{a'b'c'd'}}\left| {_{a' = c' + 1}} \right.} \right.} } } } \\
\\
\,\,\,\,\,\,\,\,\,\,\,\,\,\,\,\,\,\,\,\,\,\,\,\,\,\,\,\,\,\,\,\,\,\,\,\,\,\,\,\,\,\,\,\,\,\,\,\,\left. { + \sqrt {(a' + 1)(b' + 1)} {\rho _{a'b'c'd'}}\left| {_{a' = c' - 1}} \right.} \right).
\end{array}
\end{equation}
Since the output density operator ${\rho _{A{B_2}}}$ possesses an infinite number of elements, we are required to deploy a numerical method to approximate ${\rho _{A{B_2}}}$ by limiting its size, i.e. creating a truncated ${\rho _{A{B_2}}}$. Here, a very similar approach to that described in our previous work \cite{Neda3} can be used to estimate ${\rho _{A{B_2}}}$, and then calculate the elements of the CM ${M_{A{B_2}}}$ through the use of Eq.~\eqref{x'y'z'}. However, to determine the true resultant key rates (not just lower bounds) for the non-Gaussian state, Eq.~\eqref{output} must be utilized.

There is also an analytical approach to obtain the CM ${M _{A{B_2}}}$ \cite{nG1, nG2}, in which we first need to calculate the CM of the initial PSS state ${\rho _{A{B_1}}}$ given by
\begin{equation}\label{AB1}
\begin{array}{l}
{M_{A{B_1}}} = \left( {\begin{array}{*{20}{c}}
{xI}&{zZ}\\
{zZ}&{yI}
\end{array}} \right),\,\rm where\\
\\
x = 2v' + 1,\,\,\,\,y = 2v' - 1,\,\,\,\,z = 2\sqrt {{{v'}^2} - 1\,} ,
\end{array}
\end{equation}
and where $v' = \frac{{1 + \left( {1 + T} \right){{\sinh }^2}r}}{{1 + \left( {1 - T} \right){{\sinh }^2}r}}$. Note the PSS state ${\rho _{A{B_1}}}$ has zero mean.

After transmission of mode~$B_1$ through a quantum channel with transmissivity $\tau$ and excess noise $\varepsilon$, the CM of the resulting mixed state ${\rho _{A{B_2}}}$ is given by
\begin{equation}\label{AB2-CM}
{M_{A{B_2}}} = \left( {\begin{array}{*{20}{c}}
{x\,I}&{\sqrt {\tau \,} z\,Z}\\
{\sqrt \tau  \,z\,Z}&{\left( {\tau (y + {\chi _c})} \right)\,I}
\end{array}} \right),
\end{equation}
where ${\chi _c} = \varepsilon  + \frac{{1 - \tau }}{\tau }$. Note also the numerical method (truncated Eq.~\eqref{AB2}) gives effectively the same output CM as the CM in Eq.~\eqref{AB2-CM} for $\varepsilon=0$, provided the output density matrix ${\rho _{A{B_2}}}$ is appropriately estimated \cite{Neda3}. After mode~$B_2$ undergoes the beam splitter BS2, the CM of the resulting mixed state ${\rho _{A{B_3}}}$ is given by
\begin{equation}\label{AB3-CM}
{M_{A{B_3}}} = \left( {\begin{array}{*{20}{c}}
{x\,I}&{\sqrt {\mu \tau \,} z\,Z}\\
{\sqrt {\mu \tau \,} z\,Z}&{\left( {\mu \tau (y + \chi )} \right)\,I}
\end{array}} \right),
\end{equation}
where $\chi  = {\chi _c} + \frac{{{\chi _d}}}{\tau }$, and where ${\chi _d} = \frac{{1 - \mu }}{\mu }{\nu _d}$. Having the CMs ${M_{A{B_2}}}$ in Eq.~\eqref{AB2-CM} and ${M_{A{B_3}}}$, we are then able to compute a key rate, which will be a lower bound on the actual key rate of our CV-QKD protocol.

If the original TMSV state $\rho_{AB}$ is exploited as the source of the CV-QKD protocol, the output state at the end of the channel will still be a Gaussian state, whose CM is again given by Eq.~\eqref{AB2-CM}, but where now $x$, $y$ and $z$ are given by
\begin{equation}\label{TMSV-output-CM}
x = y = v,\,\,\,z = \sqrt {{v^2} - 1}.
\end{equation}

\subsection{Computation of the Secret Key Rate}
Here, we summarize some known results in computing a key rate based on the CM of a shared entangled state between the two trusted parties \cite{thesis, inefficient_homodyne}. It is known in the typical point-to-point CV QKD, the reverse reconciliation (RR) scenario always leads to higher key rates than the direct reconciliation (DR) scenario \cite{RR_2002}. Hence, we will only consider the RR scenario where Bob's data is the reference of reconciliation.
For a realistic reconciliation algorithm, the asymptotic key rate against collective attacks for RR is given by $K = \xi {I_{a{b_3}}} - I_E$, where $0<\xi<1$ is the reconciliation efficiency, $I_{a{b_3}}$ is the mutual information between Alice and Bob's measurements after accounting for additional detector noise and efficiencies.
$I_{E}$ is the Holevo quantity, which gives an upper bound on the quantum information stolen by Eve.

We will assume Bob makes the homodyne detection of ${{\hat q}_{{B_3}}}$, the amplitude quadrature of mode~$B_3$. On the other side, Alice invokes heterodyne detection by combining mode~$A$ with a vacuum mode on a balanced beam splitter, and applies homodyne detection on conjugate quadratures of the two output modes.
 Due to the symmetry between the two conjugate quadratures, we only consider the measurement of the amplitude quadrature of mode~$A^{he}$ (one output mode of the balanced beam splitter), which is given by ${{\hat q}_{{A^{he}}}} = \frac{1}{{\sqrt 2 }}\left( {{{\hat q}_A} + {{\hat q}_0}} \right)$, where ${{{\hat q}_A}}$ is the amplitude quadrature of mode~$A$ prior to the heterodyne detection and ${{{\hat q}_0}}$ is the amplitude quadrature of the vacuum input.
Then
 $I_{a{b_3}} = \frac{1}{2}{\log _2}\left( {\frac{{{V_{{B_3}}}}}{{{V_{{B_3}\left| {{A^{he}}} \right.}}}}} \right)$, where $V$ denotes the variance, and
\begin{equation}\label{mutual1}
{V_{{B_3}}} = \left\langle {\hat q_{{B_3}}^2} \right\rangle ,\,\,{V_{{B_3}\left| {{A^{he}}} \right.}} = \left\langle {\hat q_{{B_3}}^2} \right\rangle  - \frac{{{{\left\langle {{{\hat q}_{{B_3}}},{{\hat q}_{{A^{he}}}}} \right\rangle }^2}}}{{\left\langle {\hat q_{{A^{he}}}^2} \right\rangle }},
\end{equation}
where $\left\langle . \right\rangle $ denotes the expectation value. According to the CM ${M_{A{B_3}}}$ in Eq.~\eqref{AB3-CM} we will have
\begin{equation}\label{mutual2}
\begin{array}{l}
\left\langle {\hat q_{{B_3}}^2} \right\rangle  = \mu \tau \left( {y + \chi } \right),\\
\\
\left\langle {{{\hat q}_{{B_3}}},{{\hat q}_{{A^{he}}}}} \right\rangle  = \frac{1}{{\sqrt 2 }}\left\langle {{{\hat q}_{{B_3}}},{{\hat q}_A}} \right\rangle  = \frac{1}{{\sqrt 2 }}\sqrt {\mu \tau } z,\\
\\
\left\langle {\hat q_{{A^{he}}}^2} \right\rangle  = \frac{1}{2}\left( {\left\langle {\hat q_A^2} \right\rangle  + 1} \right) = \,\frac{1}{2}\left( {x + 1} \right).
\end{array}
\end{equation}
Thus, we find ${V_{{B_3}\left| {{A^{he}}} \right.}} = \mu \tau \left( {y - \frac{{{z^2}}}{{x + 1}} + \chi } \right)$, which according to the values of $x$, $y$ and $z$ of the PSS state in Eq.~\eqref{AB1} is given by ${V_{{B_3}\left| {{A^{he}}} \right.}} =\mu \tau (1 + \chi ) $. Hence, the mutual information for the CV-QKD protocol using the PSS states is given by $I_{a{b_3}}=\frac{1}{2}{\log _2}\left( {\frac{{y + \chi }}{{1 + \chi }}} \right)$.

Eve's information on Bob's measurement outcome, $b_3$, is given by ${I_E} = S\left( {{\rho _E}} \right) - S\left( {{\rho _{E\left| {{b_3}} \right.}}} \right)$, where $S(\rho )$ is the von Neumann entropy of the state $\rho$. To determine  $ S\left( {{\rho _E}} \right)$ we assume Eve's system $\rho _E$ purifies $\rho _{AB_2}$, that is, $S({\rho _E}) = S({\rho _{A{B_2}}})$.
 The entropy $S({\rho _{A{B_2}}})$ can be calculated through the symplectic eigenvalues ${\nu _{1,2}}$ of ${M_{A{B_2}}}$ given by Eq.~\eqref{AB2-CM}. This leads to $S({\rho _{A{B_2}}}) = f({\nu _1}) + f({\nu _2})$, where $f(x) = \frac{{x + 1}}{2}{\log _2}\left( {\frac{{x + 1}}{2}} \right) - \frac{{x - 1}}{2}{\log _2}\left( {\frac{{x - 1}}{2}} \right)$, and  $\nu _{1,2}^2 = \left( {\Delta  \pm \sqrt {{\Delta ^2} - 4\Omega } } \right)/2$, with
\begin{equation}\label{compare}
\begin{array}{l}
\Delta  = {x^2} + {\tau ^2}{\left( {y + {\chi _c}} \right)^2} - 2\tau {z^2},\\
\\
\Omega  = {\left( {x\tau \left( {y + {\chi _c}} \right) - \tau {z^2}} \right)^2}.
\end{array}
\end{equation}

The second entropy we require in order to determine $I_E$ can be written as
$S({\rho _{E\left| {b_3} \right.}}) = S\left( {{\rho _{AHG\left| {b_3} \right.}}} \right)$,
since $\rho_{AEHG}$ is pure.
 The conditional entropy $S\left( {{\rho _{AHG\left| {b_3} \right.}}} \right)$ can be calculated as $S\left( {{\rho _{AHG\left| {b_3} \right.}}} \right) = f({\nu _3}) + f({\nu _4})$, where
\begin{equation}\label{compare}
\begin{array}{l}
\nu _{3,4}^2 = \left( {\Delta ' \pm \sqrt {{{\Delta '}^2} - 4\Omega '} } \right)/2, \ \ {\rm with}\\
\\
\Delta ' = \frac{1}{{\tau \left( {y + \chi } \right)}}\left( {\tau \left( {y + {\chi _c}} \right) + x\sqrt \Omega   + {\chi _d}\Delta } \right),\\
\\
\Omega ' = \frac{{\sqrt \Omega  }}{{\tau \left( {y + \chi } \right)}}\left( {x + {\chi _d}\sqrt \Omega  } \right).
\end{array}
\end{equation}

\subsection{CV-QKD Protocol over Atmospheric Fading Channels}

In atmospheric channels the transmissivity $\eta_{tran}$ fluctuates due to several effects. Such fading channels can be characterized by a distribution of values $\eta $, with a probability density distribution $p\left( \eta  \right)$, where $\eta  = \sqrt {{\eta _{tran}}} $. Consistent with other recent studies \cite{fso, beamwander, Usenko}, we will assume that fading arising from the atmosphere is due only to beam wander. Assuming the beam spatially fluctuates around the center of the receiver's aperture, the probability density distribution $p\left( \eta  \right)$ can be described by the log-negative Weibull distribution \cite{beamwander},
\begin{equation}\
p\left( \eta  \right) = \frac{{2{L^2}}}{{\sigma _b^2\gamma_s \eta }}{\left( {2\ln \frac{{{\eta _0}}}{\eta }} \right)^{\left( {\frac{2}{\gamma_s }}- 1 \right) }}\exp \left( { - \frac{{{L^2}}}{{2\sigma _b^2}}{{\left( {2\ln \frac{{{\eta _0}}}{\eta }} \right)}^{\left( {\frac{2}{\gamma_s }} \right)}}} \right)
\label{f1}
\end{equation}
for $\eta  \in \left[ {0,\,{\eta _0}} \right]$, with $p\left( \eta  \right) = 0$ otherwise.
Here, $\sigma _b^2$ is the beam wander variance,
 $\gamma_s$ is the shape parameter,  $L$ is the scale parameter, and ${\eta _0}$ is the maximum value of $\eta$. The latter three parameters are given by
\begin{equation}\label{f2}
\begin{array}{l}
\gamma_s  = 8h\frac{{\exp \left( { - 4h} \right){I_1}\left[ {4h} \right]}}{{1 - \exp \left( { - 4h} \right){I_0}\left[ {4h} \right]}}{\left[ {\ln \left( {\frac{{2\eta _0^2}}{{1 - \exp \left( { - 4h} \right){I_0}\left[ {4h} \right]}}} \right)} \right]^{ - 1}},\\
\\
L = \beta{\left[ {\ln \left( {\frac{{2\eta _0^2}}{{1 - \exp \left( { - 4h} \right){I_0}\left[ {4h} \right]}}} \right)} \right]^{ - \left( {{1 \mathord{\left/
 {\vphantom {1 \gamma_s }} \right.
 \kern-\nulldelimiterspace} \gamma_s }} \right)}},\,\,\eta _0^2 = 1 - \exp \left( { - 2h} \right) ,
\end{array}
\end{equation}
where ${I_0}\left[ . \right]$ and ${I_1}\left[ . \right]$ are the modified Bessel functions, and where $h = {\left( {{\beta \mathord{\left/
 {\vphantom {a W}} \right.
 \kern-\nulldelimiterspace} W}} \right)^2}$, with $\beta$ being the receiver aperture radius and $W$ the beam-spot radius. In our subsequent calculations we will adopt $W=\beta$, and let the mean fading loss be controlled only by adjustments to the value of $\sigma _b$.

Since depolarization is very weak in the atmospheric channel, dephasing  will also be weak and thus we will ignore it \cite{sem}. We will also assume the transmissivity of the channel can be measured in real-time at the receiver by passing a local oscillator through the channel in an orthogonal polarized mode to the signal.

In the implementation of our CV-QKD protocol over the atmospheric fading channels towards (and from) a LEO satellite, we will assume one mode of the PSS state remains at the ground station (satellite), while the other mode (photon-subtracted mode) is transmitted to the satellite (ground station) over the fading uplink (downlink) with probability density distribution $p\left( \eta  \right)$. Let us assume Alice is the sender and Bob is receiver. After each realization of $\eta $, using the CM in Eqs.~\eqref{AB2-CM} and \eqref{AB3-CM}, where the transmissivity $\tau$ needs to be replaced by $\eta^2 $, we can compute the lower bound on the key rate given by ${K(\eta )}$. Since $\eta$ is a random variable, the elements of the final key rate is computed by averaging ${K(\eta )}$ over all possible transmission factors of the fading channel giving $K = \int_0^{{\eta _0}} {K(\eta )} p(\eta )\,d\eta $ in units of bits per pulse.

Note the key rate of the protocol must be computed by including the creation probability $P_c$ of the initial entangled state as $P_cK$. In our protocol, when the PSS state is created just-in-time from the incoming TMSV state, and then used for the CV-QKD protocol $P_c=P_{ss}$. Here, we have assumed the creation probability of the original TMSV state is one.

The range of losses we consider covers a wide range of anticipated scenarios for LEO satellite-based communications with losses in the range 0-10 dB (downlink) to 20-30 dB (uplink). Our results will also be applicable to direct line-of-sight terrestrial communications through air.

\section{Simulation Results}

\begin{figure}[!t]
    \begin{center}
   {\includegraphics[width=2.7 in, height=3.8 in]{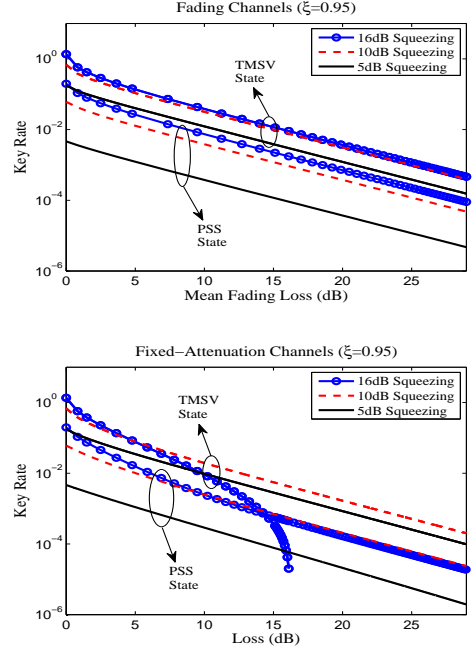}}
    \caption{The key rate (in bits per pulse) over the fading channels (top) and the corresponding fixed-attenuation channels (bottom) using the PSS state and the original TMSV state for different values of initial squeezing with $\xi=0.95$.}\label{high}
    \end{center}
\end{figure}

\begin{figure}[!t]
    \begin{center}
   {\includegraphics[width=2.7 in, height=3.8 in]{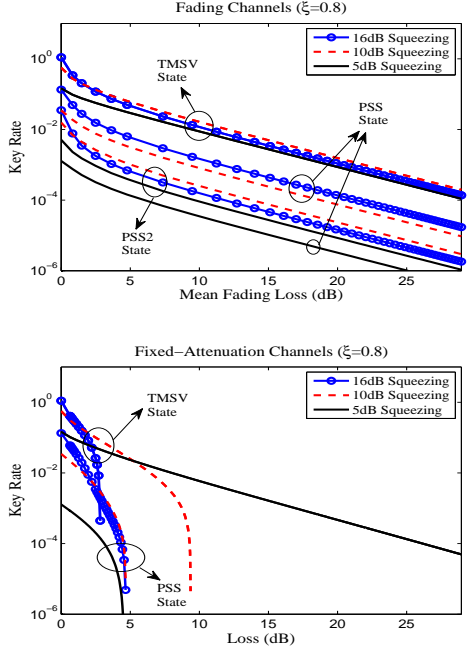}}
    \caption{The key rate (in bits per pulse) over the fading channels (top) and the corresponding fixed-attenuation channels (bottom) using the PSS state, PSS2 state, and the original TMSV state for different values of initial squeezing with $\xi=0.8$.}\label{low}
    \end{center}
\end{figure}

We now consider the performance of our CV-QKD protocol.
In Fig.~\ref{high} (top), we plot the key rate $P_cK$ resulting from the PSS states as well as the original TMSV states over the fading channels as a function of channel loss for different values of squeezing of the original TMSV state; 5dB, 10dB and 16dB.\footnote{Note for the photon subtraction operation, the transmissivity $T$ of the exploited beam splitter BS1 can be chosen arbitrarily from 0 to 1, leading to the change of the creation probability $P_{ss}$ and also the CM in Eqs.~\eqref{AB2-CM} and \eqref{AB3-CM}. Thus, for each value of loss, the key rate $P_cK$ varies with different $T$. Here, in our all simulations we have chosen an optimal value of $T$ for each value of loss to maximise the key rate $P_cK$.} As discussed earlier, in this protocol Alice and Bob apply heterodyne detection and  homodyne detection, respectively, to their own modes, followed by a RR process. First, we focus on a high-efficiency reconciliation, i.e., $\xi=0.95$. Our all simulations show the key rate in the presence of several noise sources;  the input excess noise of $\varepsilon  = 0.01$, and the efficiency  and electronic noise of the homodyne detector $\mu = 0.526$ and   $\nu_{el}=0.04361$, respectively \cite{nG1}.
The abscissa of Fig.~\ref{high} (top) corresponds to $ - 10{\log _{10}}(\int_0^{{\eta _0}} {{\eta ^2}p(\eta )\,d} \eta )$ and represents the mean fading loss in the fading channel under different conditions (different $\sigma _{b}$). The impact of a low-efficiency reconciliation is investigated in Fig.~\ref{low} (top), where we have adopted $\xi=0.8$. According to the top plots of Fig.~\ref{high} and Fig.~\ref{low}, the computed key rate resulting from the PSS state is always lower than its original TMSV state over the atmospheric fading channels.


We also simulate the performance of our CV-QKD protocol over  fixed-attenuation channels.
In order to make a valid comparison, we will assume the loss in the fixed-attenuation channel is the same as the mean fading loss in the corresponding fading channel, i.e.,  $\tau  = \int_0^{{\eta _0}} {{\eta ^2}p(\eta )\,d} \eta $. The bottom plots of Fig.~\ref{high} and Fig.~\ref{low} show the key rate  over the fixed-attenuation channels as a function of channel loss, i.e., $  - 10{\log _{10}}({\tau })$ with all the settings and parameters being the same as the corresponding protocols in the top figures.
The key rate over the fading channels is always higher than that over the corresponding fixed-attenuation channels. This fact (which comes from the probabilistic nature of the fading channels) is most evident from Fig.~\ref{low} where we see the fixed-attenuation channel is not able to generate  positive key rates (at high losses) for squeezing of 10dB and 16dB. This is not the case for the corresponding fading channels.  From the bottom plots of Fig.~\ref{high} and Fig.~\ref{low}, we can also see that for the high squeezing regime (here 16dB) which is not experimentally achievable at the moment, the PSS state can be more effective in terms of the key rates relative to the original TMSV state for some range of losses (see also \cite{nG1, nG2}).

We have also simulated the performance of the CV-QKD protocol using a state created from the incoming TMSV state by subtracting one photon from \emph{each} mode (using beam splitters with the same  transmissivity $T$).  Such a non-Gaussian state and its creation probability is given by (e.g., \cite{Oxford, Neda3})
\begin{equation}\label{PSS2}
\begin{array}{l}
\left| {PSS2} \right\rangle  = \sum\limits_{n = 0}^\infty  {{q_n}} {\left| n \right\rangle _1}{\left| n \right\rangle _2},\,\rm where\\
\\
{q_n} = \sqrt {\frac{{{{\left( {1 - {\lambda ^2}{T^2}} \right)}^3}}}{{1 + {\lambda ^2}{T^2}}}} {\left( {\lambda T} \right)^n}(n + 1),\\
\\
{P_{sb}} = \frac{{{\lambda ^2}\left( {1 - {\lambda ^2}} \right)\left( {1 + {\lambda ^2}{T^2}} \right){{\left( {1 - T} \right)}^2}}}{{{{\left( {1 - {\lambda ^2}{T^2}} \right)}^3}}}.
\end{array}
\end{equation}
The CM of the evolved $\left| {PSS2} \right\rangle$ state is again given by Eq.~\eqref{AB2-CM} but where now $x$, $y$ and $z$ are given by
\begin{equation}\label{xyz-PSS2}
x = y = \frac{{1 + 3{\lambda ^4}{T^4} + 8{\lambda ^2}{T^2}}}{{1 - {\lambda ^4}{T^4}}},\,\,\,z = \frac{{4\lambda T\left( {1 + 2{\lambda ^2}{T^2}} \right)}}{{1 - {\lambda ^4}{T^4}}}.
\end{equation}
For $\left| {PSS2} \right\rangle$ states over fading channels, we find similar trends to our previous results but  with lower key rates (at higher squeezing levels) relative to the single-photon subtracted states. However, the  point remains that these new bounds for the $\left| {PSS2} \right\rangle$  states are still less than the TMSV rates. Fig.~\ref{low} (top) illustrates this point. Over  fixed-attenuation channels the $\left| {PSS2} \right\rangle$ states showed similar trends (not plotted due to closeness of curves).

\begin{figure}[!t]
    \begin{center}
   {\includegraphics[width=2.7 in, height=3.8 in]{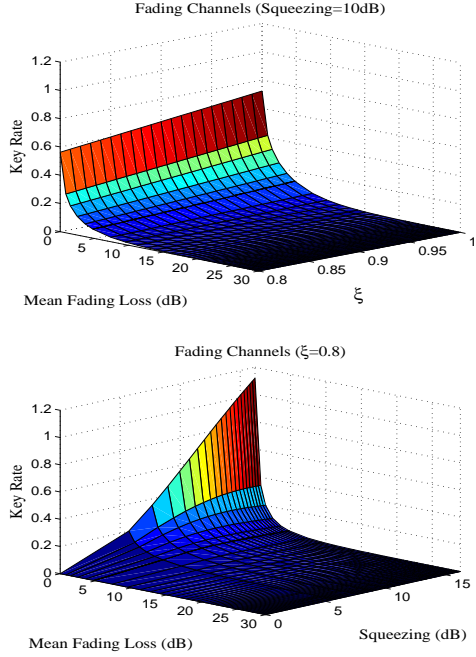}}
    \caption{The key rate (in bits per pulse) over the fading channels using the TMSV state as a function of the mean fading loss and the reconciliation efficiency $\xi$ for  squeezing 10dB (top), and also as a function of the mean fading loss and  initial squeezing for $\xi=0.8$ (bottom).}\label{3d}
    \end{center}
\end{figure}

Finally, we  investigate the impact of the imperfect reconciliation process on the resulting key rates (it will suffice to use the TMSV state for this purpose).
Fig.~\ref{3d} (top) shows the key rate $K$ of the CV-QKD protocol achieved by the TMSV state with squeezing  10dB (the state-of-the-art squeezing) as a function of the reconciliation efficiency $\xi$ and as a function of the fading channel loss. This figure clearly illustrates the negative impact of the imperfect reconciliation process on the resulting key rate. The results shown in the Fig.~\ref{3d} (bottom) illustrate the key rate $K$ with reconciliation efficiency $\xi=0.8$ as a function of the initial squeezing of the TMSV state and as a function of the fading channel loss.

According to Fig.~\ref{3d} (bottom), it is evident that an increase in fading loss reduces the key rate, while increasing the initial squeezing is able to partly compensate the fading's negative effect. However, Fig.~\ref{3d} (bottom) also shows that a rise in the initial squeezing is not always able to increase the key rate. In fact, for each value of the fading loss there is an optimal value of squeezing which maximizes the key rate. This optimal squeezing is reduced by increasing the fading loss. For instance, for a mean fading loss of $0.8$dB, the optimal squeezing is about $15$dB, while for higher loss at $11.2$dB, the optimal value is around $10$dB. Note that in case of perfect reconciliation, i.e. $\xi=1$, the key rate always grows with the initial squeezing.

Not shown in Fig.~\ref{3d} is the impact of other real-world issues such as the effect of time delays caused by the reconciliation process (and other such effects). However, the impact of the reconciliation efficiency is in most circumstances the dominant effect in reducing the quantum key rates. State-of-the-art CV reconciliation efficiencies are close to values of 0.95 \cite{eff}. At these efficiencies, achievable squeezing levels will produce key rates of $10^{-3}$ bits per pulse at mean fading loss of $25$dB.

\section{Conclusion}

In this work we have explored the usefulness of Gaussian and non-Gaussian entangled states over atmospheric channels towards LEO satellites. Focussing on the Gaussian component of both types of states as they evolve over the atmospheric channels, we have computed the quantum key rate for the Gaussian case and a lower bound for the non-Gaussian case.

We have also investigated how the presence of noise and imperfections of the protocol affect the key rates. Relative to the corresponding fixed-attenuation channels (such as those found in fiber), we find the fading channels (due to their stochastic nature) can lead to higher key rates. This somewhat counter-intuitive result is most noticeable for the very high-loss channels. We have also found that in some circumstances the non-Gaussian states could result in enhanced quantum key rates over the fixed-attenuation channels compared to the Gaussian states. This is the opposite of what we find in the atmospheric fading channels, once again highlighting the impact the atmosphere can have on the quantum key outcomes. We caution that our rates for the non-Gaussian states are lower bounds - the true key rate for such states remains an open problem. If the bounds shown are here not tight the relative performance levels discussed here may vary.

The role played by the efficiency of the classical reconciliation process, as a function of the mean fading loss and the initial squeezing of the Gaussian states, was also investigated. For state-of-the-art squeezing levels (10dB) and state-of-the-art reconciliation efficiencies (0.95) we found ground-to-satellite quantum key rates of roughly $10^{-3}$ bits per pulse are viable. This channel will be the limiting channel in a ground-satellite-ground relay system. Due to the reduced beam wandering losses incurred from space to Earth, the key rates of the satellite-to-ground channels will be roughly an order of magnitude larger.

The calculations presented here should be of value in the quantitative assessment of the CV-QKD protocols using Gaussian and non-Gaussian resources for future quantum communications over atmospheric channels.


\end{document}